\documentclass{osa-article}

\usepackage{amsmath}

\usepackage{multirow}
\usepackage{lipsum}
\usepackage[export]{adjustbox}

\newcommand{\donotshow}[1]{}

\begin{document}

\title{Artefact removal in ground truth and noise model deficient sub-cellular nanoscopy images using auto-encoder deep learning}

\author{Suyog Jadhav\authormark{1}, Sebastian Acuna\authormark{2}, Krishna Agarwal\authormark{2}, and Dilip K. Prasad\authormark{3,*}}

\address{\authormark{1}Indian Institute of Technology (Indian School of Mines), Dhanbad 826004, India\\
\authormark{2}Department of Physics and Technology, UiT The Arctic University of Norway, Troms\o, Norway\\
\authormark{3}Department of Computer Science, UiT The Arctic University of Norway, Troms\o, Norway}

\email{\authormark{*}dilip.prasad@uit.no} 



\begin{abstract}
Image denoising or artefact removal using deep learning is possible in the availability of supervised training dataset acquired in real experiments or synthesized using known noise models. Neither of the conditions can be fulfilled for nanoscopy (super-resolution optical microscopy) images that are generated from microscopy videos through statistical analysis techniques. Due to several physical constraints, supervised dataset cannot be measured. Due to non-linear spatio-temporal mixing of data and valuable statistics of fluctuations from fluorescent molecules which compete with noise statistics, noise or artefact models in nanoscopy images cannot be explicitly learnt. Therefore, such problem poses unprecedented challenges to deep learning. Here, we propose a robust and versatile simulation-supervised training approach of deep learning auto-encoder architectures for the highly challenging nanoscopy images of sub-cellular structures inside biological samples. We show the proof of concept for one nanoscopy method and investigate the scope of generalizability across structures, noise models, and nanoscopy algorithms not included during simulation-supervised training. We also investigate a variety of loss functions and learning models and discuss the limitation of existing performance metrics for nanoscopy images. We generate valuable insights for this highly challenging and unsolved problem in nanoscopy, and  set the foundation for application of deep learning problems in nanoscopy for life sciences. %
%
%
%
\end{abstract}

\section{Introduction}

\begin{figure}[t]
    \centering
    \small
    \includegraphics[width=0.8\linewidth]{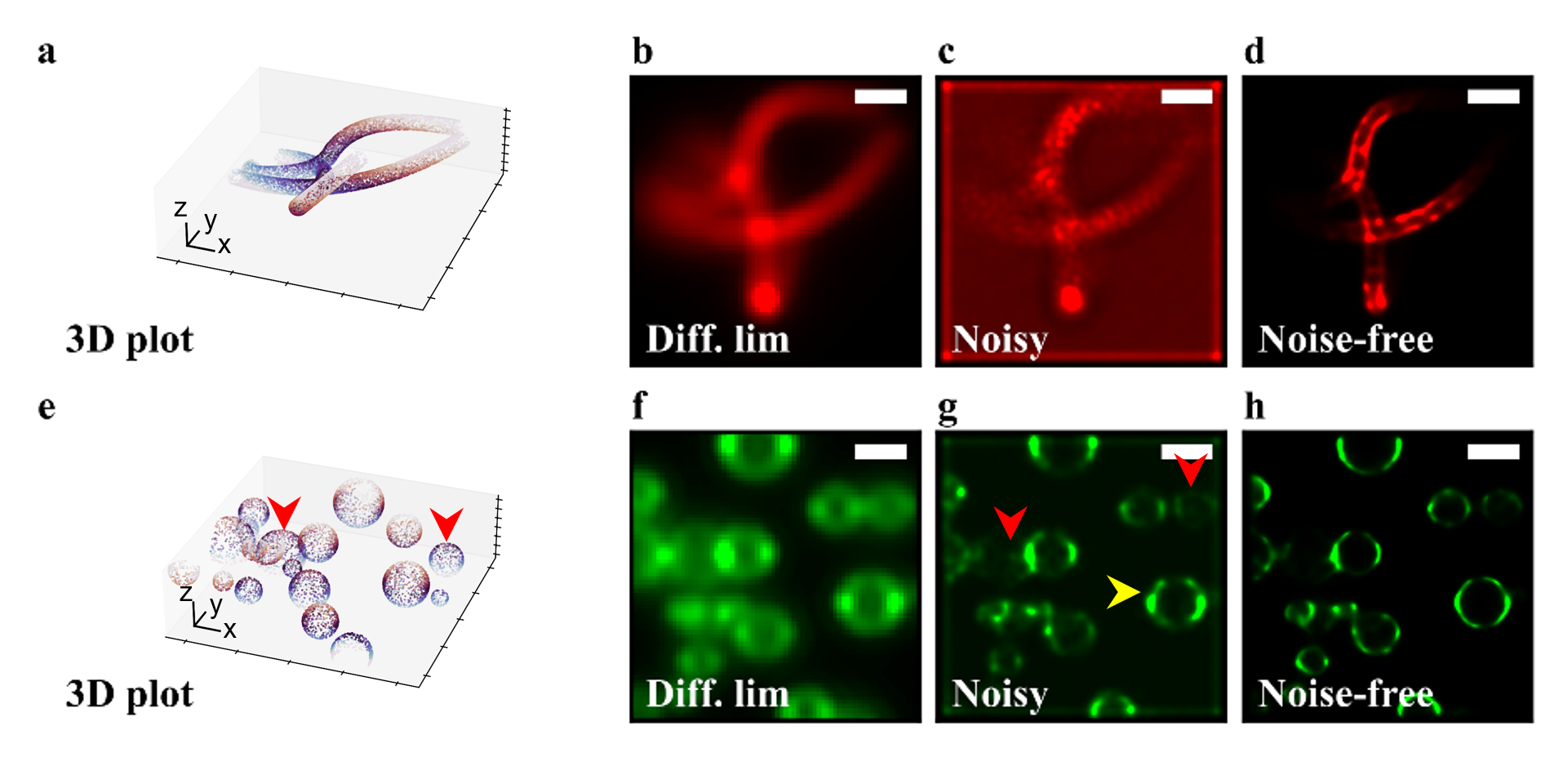}
    \caption{A side by side view of noisy (SBR3) and ideal MUSICAL reconstructions (simulated) is presented.The top row shows an example of mitochondria while the bottom row shows an example of vesicles. In some case, as shown in (c,d) artefacts can suppress resolvability of features in addition to contributing background debris. In other cases, it may compromise the sharpness of certain structures (yellow arrows in g) and reduce optical sectioning by reconstructing out-of-focus structures (as shown with red arrows in g). Number of frames used to generate nanoscopy images 200. Diffraction limited image (mean image of all frames) is abbreviated as Diff. Lim.  Scale bar 500 nm in 3D plot and 1 $\upmu$m. in [b-d] and [f-h].}
    \label{fig:ideal-noisy}
\end{figure}

This article addresses the problem of artefact removal, a form of denoising problem, in nanoscopy (i.e. super-resolved optical microscopy) images obtained using computational nanoscopy techniques \cite{dertinger2009fast, gustafsson2016fast,agarwal2016multiple,doi:10.1021/acsphotonics.5b00307,Hu2013,10.1371/journal.pone.0094807,Zhao2018FasterSI, Solomon:18} used for sub-cellular imaging of biological cells. These techniques take in a high speed fluorescence microscopy video (also called the raw microscopy data or raw microscopy image stack) comprising of 10s to 100s frames, which are fast enough to capture the fluctuations arising from photokinetic nature of fluorescent emitters (referred to as emitters for simplicity) \cite{dempsey2011evaluation,rollins2015stochastic} used to label a sample, and perform statistical analysis of these fluctuations to construct super-resolved images. We refer to these methods as fluctuations based nanoscopy methods (FNMs). As reported in \cite{2008.09195}, artefacts appear to be an unavoidable feature of FNMs because of the nonlinear statistical analysis tools used as the backbone. Fig. \ref{fig:ideal-noisy} shows examples of two different sub-cellular structures, the average image of the raw stack (referred to as diffraction limited image), and the nanoscopy images generated by multiple signal classification algorithm (MUSICAL \cite{agarwal2016multiple}, an example FNM) using noisy and noise-free raw microscopy data. Being less phototoxic and more live cell compatible, FNMs are highly desirable for conducting nanoscale studies in life sciences but the artefacts in nanoscopy images can interfere in deriving correct inferences. Therefore, suppressing artefacts in them is an endeavour of significant impact. 

Artefact removal in the nanoscopy images can be considered as a version of denoising problem in the sense that artefacts are associated to the noise characteristics and have to be removed from the image similar to the need of removal of noise in denoising.
An important deviation must be noted though.
Artefacts may not be completely stochastic in nature as opposed to the general noise distributions since they encode the stochastic parameters of noise and photokinetics as well as the systematic distortion introduced by the microscope or algorithm. 
Nonetheless, for simplicity of reference, we use the terms noisy, noise-free, and denoising for artefact-ridden, artefact-free, and artefact removal respectively. 
Also, unless specified otherwise, these terms apply in the context of the processed nanoscopy images rather than the raw microscopy image stacks used for generating the nanoscopy images.

Deep learning based denoising of signals and images has gained quite some traction in the recent times \cite{burger2012image,zhang2017beyond,plotz2017benchmarking,guo2019toward,brooks2019unprocessing,singh2020single}, even for denoising microscopy images \cite{Wang2019}. Assumption of large supervised training dataset is an inherent assumption in deep learning, which is often difficult to achieve since generating pair of noisy and noise-free images using the same sensor is not possible. Therefore, the noise model is assumed to be known and is used to create synthetic supervised training datasets. Additive white Gaussian noise is often assumed \cite{burger2012image,zhang2017beyond}, which provided state-of-the-art denoising performance when released. However, another contemporary study indicated that traditional methods work better in most real scenarios of noisy images \cite{plotz2017benchmarking}. This is a prime reason for using traditional approaches such as feature based reconstruction for microscopy images even in recent times \cite{Haider2016,liu2017scmos,meiniel2018denoising,Maji2019,mandracchia2020fast}. This led to the appreciation that synthesizing the right noise model is a key to quality denoising \cite{guo2019toward,brooks2019unprocessing,meiniel2018denoising}. 

In microscopy data, there are two major sources of stochastic noise, namely the shot noise (Poisson distribution) arising from the photon scattering behaviour and the electronic noises of cameras, whose noise models depend on the type of scientific cameras \cite{liu2017scmos,mandracchia2020fast}. There are other systematic sources of artefacts, such as camera drift, microscope aberrations, and occasionally dead and hot pixels. Systematic artefacts due to these sources can be greatly reduced or even completely removed by changing or upgrading the microscope hardware. In some cases, the effect of electronic noises and shot noise can be reduced by using extremely high light doses for non-fluorescent microscopes, such as used for creating a supervised microscopy dataset in \cite{Manifold:s}. 

However, neither creating a supervised training dataset nor modeling the noise or artefacts is an option in FNMs due to multiple reasons, as explained next. For creating an experimentally measured supervised training dataset, a pair of identical raw microscopy data should be measured, except that one is noisy while the other is noise-free. This is possible to some extent for individual fluorescence microscopy images \cite{Nguyen2018DeepLF,zhang2019poisson}, but impossible for videos due to the following:
\begin{itemize}\itemsep0em
    \item \textbf{Fluorescence bleaching effect:} Unlike \cite{Manifold:s,Nguyen2018DeepLF} where a high dose of light was used to generate low noise images for emulating noise-free raw data, using high light dose for getting the equivalent raw noise-free videos bleaches the fluorescent samples and instead introduces fast decreasing intensity effect which does not match the low light dose image. 
     \item \textbf{Impossible to replicate the fluctuation statistics:} The emission of photons from fluorescent molecules is a stochastic process. Therefore, it is impossible to replicate the temporally precise series of emissions between the two set of measurements. Further, the averaging approach over multiple frames considered in \cite{zhang2019poisson} cannot be used since it modifies the manifestation of fluctuations in the averaged stack and also does not match the temporal rate of noisy image stack. 
\end{itemize}
Further, generating a noise or artefact model is not possible for FNMs. Each image in the raw microscopy image stack itself is a linear map of the microscope's transfer function (conventionally called the point spread function in microscopy) convolved with the emitter locations and weighted by the number of photons emitted by these emitters during that frame as a consequence of their photokinetics. The point spread function (PSF) and the image characteristics encode optical properties such as numerical aperture, wavelength of fluorescence, and camera pixel size. Known camera-specific noise models also apply to individual images. But, as FNMs perform spatio-temporal mixing through statistical analysis and generate non-linear functions that indicate the presence of emitters, the fluctuations, noise, and microscope parameters all get non-linearly mapped into the nanoscopy images. These mappings have a strong dependence upon spatio-temporal density of photon emissions, light dose and its temporally non-linear effect on both photokinetics and noise level, the PSF, the statistical techniques, and the control parameters used for the FNMs, for example as discussed for super-resolution optical fluctuations imaging (SOFI) method \cite{yi2020cusp}, often resulting into custom artefacts of complicated nature. For example, in Fig. \ref{fig:ideal-noisy}, the artefacts for the mitochondria example appear in the form of background debris and compromise in resolvability of the foreground structures, while the artefacts in the vesicles example are in the form of blur edges, loss of certain details, and non-negligible visibility of out-of-focus structures. Furthermore, a wide range of fluctuation statistics, density of emitters, and microscopes may be encountered in practice, making it difficult to learn a generally valid artefact model for the chosen FNM. Therefore, learning or generating noise models is also not practically feasible.  
In brief, this particular artefact removal problem is not only ground truth deficient but also deficient of noise or artefact model. Therefore, there is no appropriate guide for creating experimental or synthetic supervised training dataset in the conventional sense. At the same time, the complex nature of artefacts indicate the need of black box deep learning approaches such as autoencoders \cite{vincent2010stacked} which mandates supervised training datasets. One potential solution is to first denoise the individual microscopy images using a suitable microscopy image denoising approach and then pass the denoised raw microscopy data to FNM. However, such an operation introduces non-linear computational distortion in the raw microscopy data, both to the noise components and the fluorescence fluctuations component. This distortion renders denoised microscopy image stack as unsuitable for FNMs. 

In view of these compound challenges, we take an unconventional route towards deep learning, namely simulation-supervised training dataset. We create the pair of noisy and noise-free nanoscopy images through simulating two exactly identical raw microscopy stacks, except that one is passed through a noise engine and the other is not. A key to the success of simulation supervised deep learning is the ability of the simulation engine to simulate the real scenarios and the corresponding ground truth. It becomes further paramount in problems that have implications on scientific inference. Therefore, each physical phenomenon underlying the raw microscopy video generation has been simulated to the needed accuracy. Example includes simulating practical range of photokinetics and even including the glass coverslip used to cover the sample in our simulation engine. In order to obtain an artefact removal approach that is robust and versatile across a wide variety of situations, structures, and microscope parameters, we consider a range of simulation parameters. 
We note that simulation supervised approaches is not necessarily new in microscopy domain \cite{sekh2020simulationsupervised,yao2020machine,gupta2020cryogan}, but the development of simulation engines that can create customized problem and microscopy modality specific as well physically loyal data is a fairly recent practice. Specifically, a simulation-supervised deep learning approach for `nanoscopy' denoising for sub-cellular structures is used for the first time. Here, we demonstrate artefact removal for one candidate FNM, namely multiple signal classification algorithm (MUSICAL) \cite{agarwal2016multiple}, although the concept is generalizable to any FNM. Beside showing versatility to denoising images with sub-cellular structures and structural density for which training has not been performed, we also show that the method is robust to noise models not simulated in the raw microscopy data. We attribute this robustness to both the diversity of other conditions in the simulated dataset as well as the fact that the artefacts created by FNMs are related to the noise and other stochastic parameters in a complex manner such that the manifestation of only the distribution of noise cannot be singled out. The highlight of our results is the quality of artefact removal on actual experimental public data of sub-cellular structures, including in fixed and living cells. 

The outline of the paper is as follows. Section 2 presents the proposed method while section 3 presents diverse validation and experimental results. Discussion and insights are presented in section 4. Section 5 concludes the work. 

\begin{figure}[t]
    \centering
    \includegraphics[width=\linewidth]{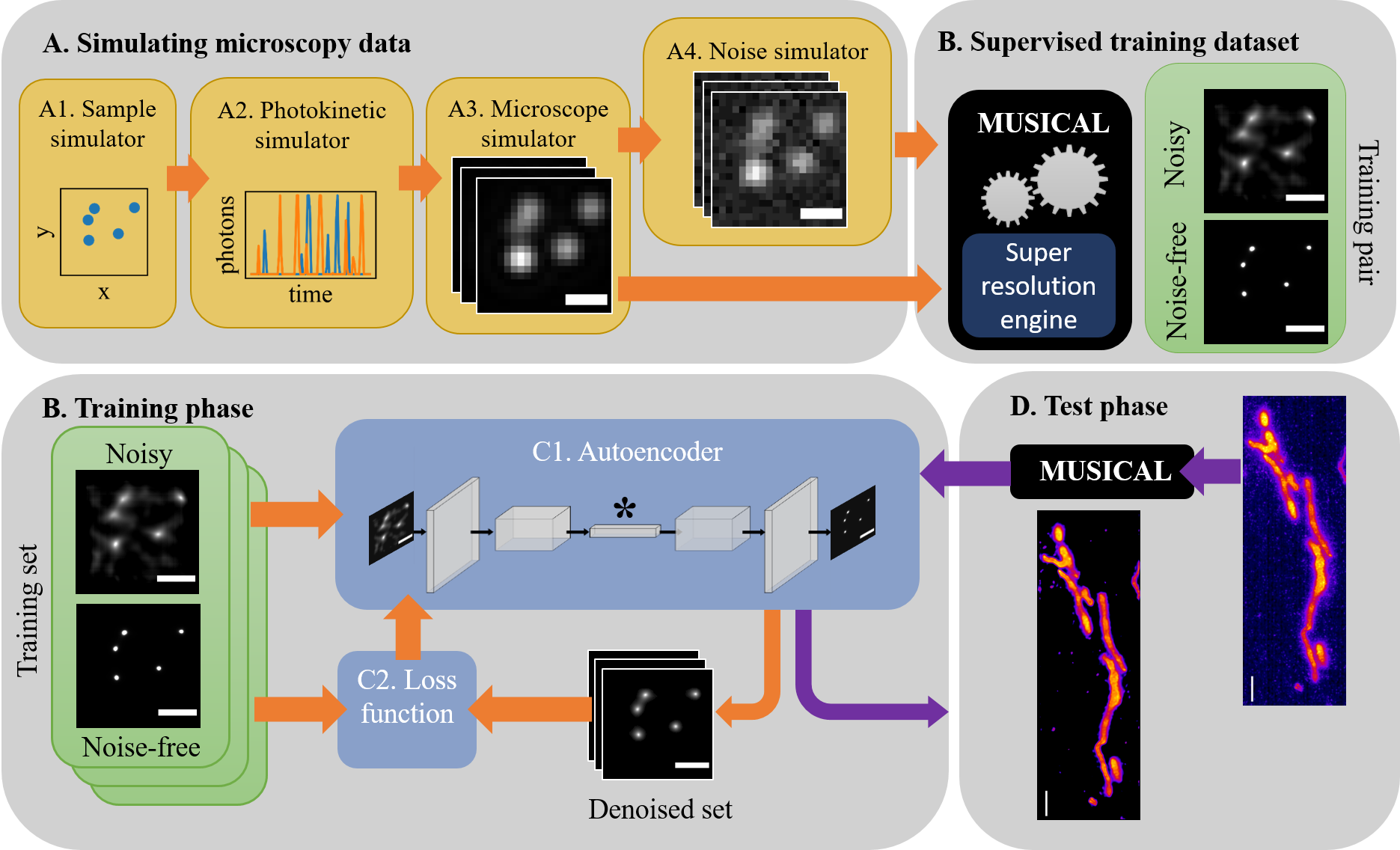}
    \caption{The proposed approach of artefact removal is illustrated here. The asterisk (*) shown in block C1 indicated the low-dimensional latent feature space of autoencoder, suitable for representing feature-deficient microscopy and nanoscopy images. Relevant details of the labeled blocks appear in section \ref{sec:proposed}.}
    \label{fig:proposed}
\end{figure}

\section{Proposed approach}\label{sec:proposed}

Our approach is shown in Fig. \ref{fig:proposed}. We first create a simulated training dataset as illustrated in block A and B of the \ref{fig:proposed}. For this first two sets of raw microscopy image stacks are simulated using precisely the same physical characteristics, with the exception that one raw microscopy image stack is noisy since it is generated by the noise-free raw microscopy image stack passing through a noise simulator. Both the raw microscopy image stacks are individually processed using MUSICAL to obtain corresponding noisy and noise-free nanoscopy images. Since the simulated dataset is used for training, it is imperative for the simulated dataset to emulate the relevant aspects of reality as closely as possible while retaining enough diversity across the simulated conditions. Several thousands of such pairs are generated and used as the supervised training dataset for the autoencoder. Then, through a good choice of autoencoder architecture and the loss functions, the autoencoder is trained for denoising the nanoscopy images. In the test phase or actual field use, raw microscopy image stack obtained by a real microscopy experiment is processed through the MUSICAL algorithm to obtain a noisy nanoscopy image, which is then passed through the trained autoencoder to generate the corresponding noise-free nanoscopy image. We discuss the details of the various blocks in the subsequent sub-sections. 

\subsection{Raw microscopy data simulator (Blocks A1-A4 of Fig. \ref{fig:proposed})} \label{sec:raw}

\paragraph{A1: Sample simulator $-$} 
The concept is that the shape and size hypotheses created by prior studies are used to simulate sample geometries. In this work, we consider three types of sub-cellular structures, namely actin filaments \cite{chiu1999high,galkin2010structural,egelman1982f}, mitochondria \cite{fawcett1966atlas,stephan2019live,rafelski2013mitochondrial}, and vesicles \cite{huang2017formation,de2020lysosomal,huotari2011endosome}. However, the setup is easily scalable to include other types of sub-cellular structures. First, the 3D geometries of the structures are simulated. Then, the positions of fluorescent molecules (called emitters for simplicity) are stochastically generated as labeling the structures. 

For simulating an actin filament, a 3D smooth curve is created by selecting certain number of spline control points and then fitting a spline through them. 
The number of control points is also selected randomly from the range $[3,6]$. 
The maximum length allowed for a filament is kept at 5 $\upmu$m. The emitters are placed randomly across the length of the spline curve with linear density of 100 emitters/$\upmu$m. This is based on two assumptions. First, the periodicity of binding sites in actin is 5-7 nm. Second, the labeling efficiency is never 100\%. Assuming 30-50\% labeling efficiency, the selected emitter density is reasonable. 

For simulating a single mitochondrion, first the spline similar to actin filament is considered. Then, a curvilinear cylinder of radius 150 nm is fit over it by convolving a cylinders of the chosen radius and height 1 nm over the spline. The selection of the diameter 300 nm is close to diffraction limit of most microscopes and its outer membrane label is not distinguishable in raw microscopy data with noise, but is expected to be reconstructed as outer boundary by a nanoscopy method. Further, as seen in Fig. \ref{fig:ideal-noisy}, under significant noise, the membrane boundary may not be explicitly reconstructed. So, we consider this radius as a border line situation of failure of MUSICAL under noise. However, other ranges of diameters may be included in the future. After constructing the geometry, the emitters are distributed randomly on the surface of the geometry with an emitter surface density 500 emitters/$\upmu$m$^2$. This emulates outer membrane label of mitochondria. The emitter density is chosen heuristically based on expert input.

A vesicle is simulated as a sphere of radius randomly chosen from the range [25,500] nm. The emitters are distributed on its surface with an emitter density of 2000 emitters/$\upmu$m$^2$, chosen heuristically. The surface labeling emulates the membrane of vesicles. 

There may be multiple instances of a structure in an image region, however only one type of structure is expected in one fluorescent color channel. Therefore, we simulate multiple actin filaments, multiple mitochondria, or multiple vesicles in each example. The number of them in a single image is chosen randomly from the range [3,10], [1,4], and [10,30] for actin filaments, mitochondria, and vesicles respectively. We impose some boundaries on the 3D space in which the sample may be present. These are $x,y \in \left[-2.5,2.5\right] \upmu$m and $z\in [-500,500]$nm where $z=0$ represents the focal plane of the microscope. 

\paragraph{A2: Photokinetics simulator $-$}
In reality, there are multiple distributions associated with the emissions of photons when the fluorescent molecule is active, the fluorescent molecule entering, dwelling, and exiting the dark states, the photobleaching etc. \cite{ doi:10.1146/annurev-physchem-032210-103340, Dickson1997, Cox2012}. 
However, at image acquisition rates of milliseconds to seconds, the need of knowing and simulating individual distributions is obviated, and simpler probability distributions can be used to represent the macro-behaviour of fluctuations in photon emissions arising from photokinetics. This simplification may not apply if specific dyes are used with long dark states, but this is neither the requirement of fluctuations based nanoscopy techniques nor are the regime in which they provide a particular advantage over other localization based methods \cite{rust2006sub,Schnitzbauer2017}.

Therefore, we use the simpler photokinetic model based on the implementation of \cite{10.1371/journal.pone.0161602}. 
In this model, a single emitter is characterized with a 2-state model.
The states are simply called \textit{on} in which the molecule is producing photons, and \textit{off} in which case no photons are emitted.
The time the emitter stays in each state is modeled with an exponential distribution controlled by two parameters called $\tau_{\rm on}$ and $\tau_{\rm off}$.
These correspond to the mean time the emitter spends in each state.
The emission rate of photons is considered constant and therefore, the number of photons emitted while the emitter is in the \textit{on} state is just the rate by the total time.
As a result, the duty cycle is then $\tau_{\rm on} / (\tau_{\rm on} + \tau_{\rm off})$. All emitters are considered identical and therefore all of them in a sample have the values of $\tau_{\rm on}$ and $\tau_{\rm off}$.

In order to emulate a range of photokinetic behaviour, we choose the values of $\tau_{\rm on}$ and $\tau_{\rm off}$ as integers taken from the ranges $[1,5]$ and $[1,20]$, respectively. It is of interest to observer that the pair $(\tau_{\rm on}, \tau_{\rm off})$ having a value (5,1) indicates extremely dense fluctuations, i.e. an extremely challenging condition for fluctuation based nanoscopy techniques where they do not provide significant resolution enhancement. On the other hand the pair having value $(1,20)$ is a conducive regime for such techniques. 

\paragraph{A3: Microscope simulator $-$}
The imaging function of the microscope is simulated using Gison Lanni model of point spread function (PSF) \cite{Gibson:92}. We use a fast implementation of Gibson Lanni PSF reported in \cite{Li:17}. The PSF simulates the blurring introduced by the optics of the microscope as the light passes through the coverslip and microscope optics to the image region where the camera is placed. 

Among the various parameters needed for simulating the Gibson Lanni PSF, the following were used as a constant for the setup. The sample is assumed to be mounted on a glass surface (such as slide) and present in water medium. A glass coverslip of 170 $\upmu$m is assumed to be present between the sample and the microscope optics. The numerical aperture (NA) of the system is selected randomly from the range $[1.2, 1.49]$. For simulation purposes, the emission wavelength of the emitters is assumed to 660 nm. In practice, the emission wavelength is a characteristic of the fluorescent dye chosen for the experiment for a particular type of structure and is generally in the range $[488,650]$ nm for visible range fluorescent dyes. However, the manifestation of the wavelength is in terms of achievable resolution and the spread of PSF. The same effect can be achieved through varying the NA of the microscope. Therefore, choosing a fixed wavelength but sufficiently large span of NA allows us to simultaneously consider variety of microscopes and dyes without loss of generalization. Since the PSF is computed in the image region to construct the image of an emitter, the camera's pixel size is also needed. The camera pixel size in terms of the sample dimensions is computed by dividing the actual hardware pixel size of the camera with the magnification of the microscopes. We consider pixel sizes in sample dimensions directly and select candidate values most popularly encountered in high NA microscopy systems. Four different pixel sizes were considered for simulation (65, 80, 108 and 120 nm), each pixel size getting used for exactly one quarter of the total number of samples simulated for each type of structure. 

\paragraph{A4: Noise simulator $-$}
The noise simulation approach is taken from \cite{agarwal2016multiple,arif2020cvpr}. There are two main sources of noise. The first is the camera's electronic noise that contributes a noisy background in the image. The second is photon noise, which is based on Poisson statistics of arrival of photon at the expected location.
Let the simulated microscopy image, scaled to span $[0,1]$ be denoted as $\bf I$. Moreover, let the signal to noise ratio be $\rm SNR$ and the measured background values in the camera with closed shutter be $b$. First, a microscopy image of the expected signal strength (such as observed in the microscopy data) and having a constant background $b$ is simulated as $\hat{\bf I} = b (\textrm{SNR}-1) {\bf{I}}  + b$. 
Then, the noisy microscopy image $\tilde { \bf I}$ is generated such that each pixel in $\tilde {\bf I}$ is generated using a Poisson distribution with mean equal to the corresponding pixel in $\hat {\bf I}$. 

With $\sim$ms exposure time used in FNMs, the electronic noise is significantly stronger than the photon noise. In such situation, signal to background ratio (SBR) is a practical measure of noise. The original article of MUSICAL reports super-resolution for SBR $\geq$ 3. Therefore, we simulate our dataset with the lowest SBR (i.e. highest level of noise) recommended for MUSICAL. Furthermore, we noted that a large variety cameras have background noise in the range $[50.120]$ on a 16 bit intensity scale, depending upon the type of camera, the imaging speed, the cooling system, and other usage factors. We used a constant value $b=100$ in our simulations. 

A total of 3000 noise-free and noisy image stacks were created, each containing 200 frames. Among them, 1000 pairs simulated each for actin filaments, microtubules and vesicles. 75\% of the pairs were used for training and 25\% were used for testing. The selection was performed randomly.

\subsection{MUSICAL (Block B of Fig. \ref{fig:proposed})}

For each pair of raw microscopy data, MUSICAL is applied independently on the noise-free and nanoscopy raw microscopy image stacks to obtain one pair of training data. Here, we explain MUSICAL and the MUSICAL parameters.

MUSICAL achieves super-resolution by performing spatio-temporal analysis of the fluctuations in the measured image stack and exploiting that the noise is stochastic while the fluctuations arising from photokinetics are modulated through the PSF in the microscopy images.
MUSICAL decomposes the image stack using singular value decomposition or eigenvalue decomposition \cite{agarwal2016multiple, agarwal2017eigen, Acuna:Thesis:2019} into a orthogonal set of vectors called eigenimages, and eigenvalues uniquely associated to them.
In particular, the eigenimages with high eigenvalues are associated to the actual emitters and therefore, are strongly related to the PSF of the system. Specifically, eigenimages associated to the actual structure are expected to be linear combinations of the PSFs at emitter locations. 
These eigenimages (the ones with high eigenvalues) are grouped into one set, called the signal subspace, since they span the images measured in the stack.
Notably, only a subset of all the eigenimages belongs to this set.
The ones that do not, are grouped into another set called the noise subspace.
The key property exploited by MUSICAL is as follows. The signal and noise subspaces are orthogonal, and the signal subspace is given by the linear combinations of PSFs at the emitters. Therefore, the PSFs at emitter locations are also orthogonal to the noise subspace.
As a result, a test point at an emitter location, will have a large projection in the signal subspace and small in the noise subspace.
On the other hand, if a test point is far from an actual structure, it has small projection in the signal subspace, and large projection in the noise subspace.
These two situations are combined in a so-called `indicator function' that takes the ratio of the projection in the signal and noise subspace.
As a result, the function is high for test points at emitters locations and low otherwise.

MUSICAL needs the following knowledge about the microscopy data: the emission wavelength of the fluorophore (or equivalently the collection wavelength of the microscope), the pixel size of the camera as scaled for sample dimensions, and the NA of the microscope. In addition, MUSICAL needs three algorithmic control parameters: (a) a threshold for assigning eigenimages to the signal and noise subspace, (b) a contrast parameter $\alpha$, and (c) the level of subpixelation which determines the fineness of the grid and pixel size in the nanoscopy image. We used a recent work on automatic soft threshold for the first parameter, which obviates the need for user-specified threshold \cite{}. Further, $\alpha$ has been set to 4 following the recommendation of \cite{agarwal2016multiple}, and subpixelation of 10 since this subpixelation gives pixel size well below the smallest structures we have considered.

\subsection{Training for denoising (Block C of Fig. \ref{fig:proposed})}
\paragraph{Autoencoder architectures}
As noted in \cite{arif2020cvpr,parnamaa2017accurate,hay2018performance}, microscopy and nanoscopy data poses several challenges as compared to the normal computer vision data because of absence of color, texture, and edge features. However, the small latent space of autoencoders (such as shown in block C1 of Fig. \ref{fig:proposed}) is an efficient way of exploiting the sparsity and lack of feature variety which is characteristic of microscopy and nanoscopy images.

\begin{figure}
    \centering
    \begin{tabular}{cc}
        \includegraphics[width=0.7\linewidth]{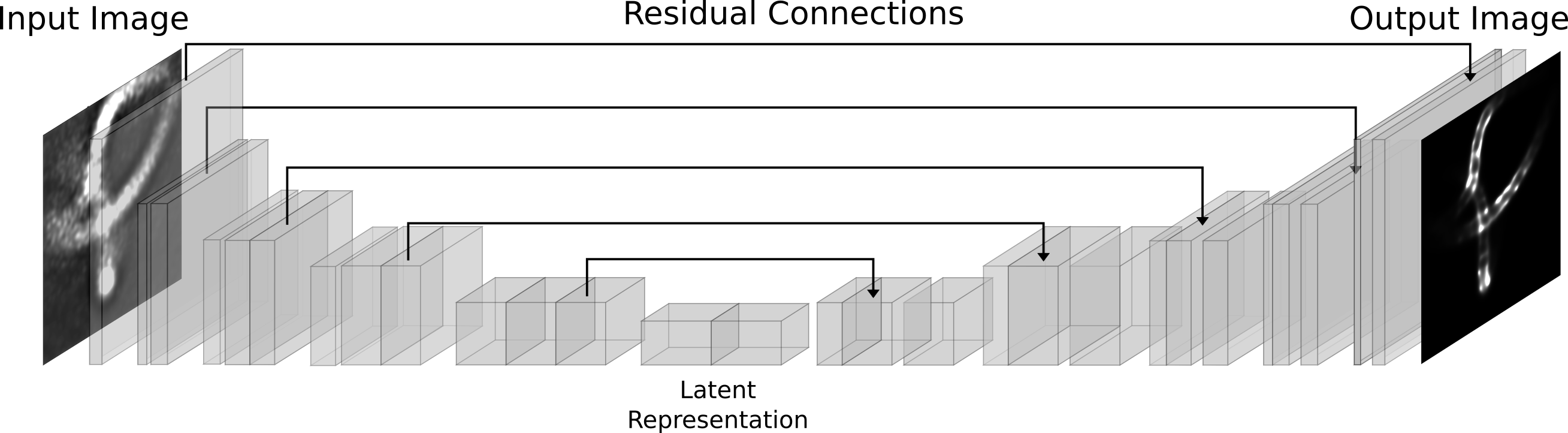} \\ \small{(a) UNet} \\ \includegraphics[width=0.7\linewidth]{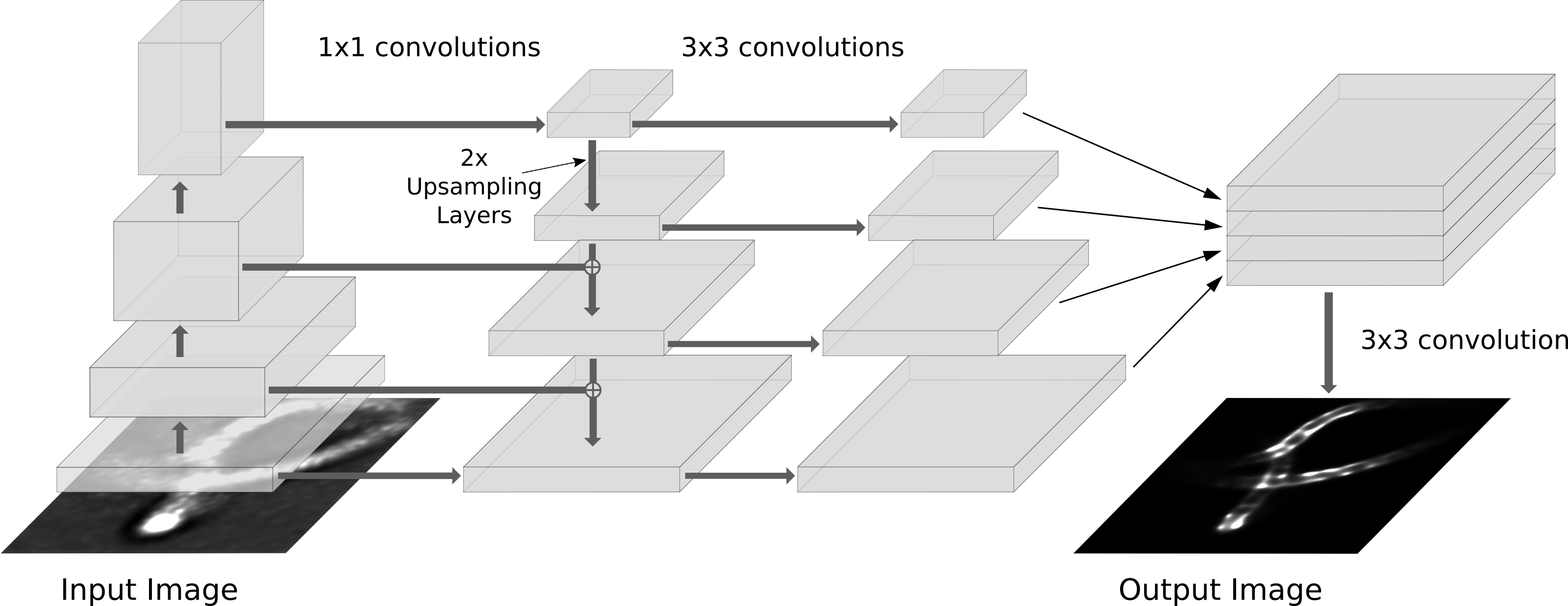}  \\ \small{(b) Feature} pyramid network (FPN)
    \end{tabular}
    \caption{Block diagrams of the autoencoder architectures explored in this work.}
    \label{fig:architectures}
\end{figure}

We tried two different architectures for this task. 
The first one is the U-Net \cite{ronneberger2015u} model. 
It was designed specifically for biomedical images and it is known for good performance in the medical imaging domain. 
The inspiration behind using U-Net is primarily the similarity of the application domain.
The second model is the Feature Pyramid Network (FPN) \cite{lin2017feature}. Although FPN was originally designed for object detection tasks, many have successfully utilised the architecture for image-to-image tasks like semantic segmentation and instance segmentation \cite{seferbekov2018feature,kirillov2019panoptic}. Inspiration for this choice was to see if the impressive performance seen with an image-to-image task like segmentation can transfer to a denoising task like ours. The architectures are shown in Fig. \ref{fig:architectures}. For each model, we considered two options for convolutional layer architectures, namely R-34 and R-50, where R stands for residual network. This was done to explore both deep and deeper architectures. We found that FPN with R-50 model often did not converge while training. So, we drop discussion on this combination hereon.

We note that some changes in the input images and the architectures were made to accommodate for the special case of the chosen nanoscopy algorithm, as described next. 
The simulated input images had 32-bit floating point pixel values. Both the input and output images were normalized using max normalization. Without the normalization, the neural network has to deal with an ill-defined problem as the actual dynamic range of the data may be much smaller than 32 bit for the noisy nanoscopy image. This is a consequence of the MUSICAL's nanoscopy performing indicator function. Further, learning  the intensity span of the actual 32 bit image for the noise-free nanoscopy for each case is more challenging than defining the intensity in the output image to be in the range $[0,1]$. Therefore, the max normalization makes the input and output intensity ranges better-defined and mapping more learnable. At the same time, loosing the actual intensity value in the output is not considered a problem since the quality of the output image is unaffected and its interpretability unaltered due to it. This is because MUSICAL and several other FMNs are qualitative reconstruction techniques in the sense that the intensity values generated by them indicate statistical significance of presence of emitters but not values of physical quantities. 
The only exception to best of our knowledge is balanced super-resolution optical fluctuation imaging (b-SOFI) \cite{Geissbuehler2012}. The selected architectures is then modified to fit the new output format of 32 bit floating point images with intensity range $[0,1]$. To do so, a rectified linear unit (ReLU) activation layer is added at the output to force the lower limit of the output image to be greater than zero. This layer is followed by a max normalization step to limit the intensity values in the output image between 0 and 1.

\paragraph{Choice of the Loss Function} The choice of the loss function determines the nature and quality of learning. 
Since nanoscopy image denoising for FNMs is new, we experimented with a variety of loss functions presented below. 
We use the following notations. The input denoised image (the output of the autoencoder) is denoted as $\hat{\bf{I}}$ while the corresponding noise-free image (the target or ground truth for the autoencoder) is denoted by $\bf{I}$. The pixel indices are specified by $n$ and the total number of pixels is $N$. Therefore intensity in the denoised image for $n$th pixel is denoted as $\hat{\bf I}_n$ and similarly for the noise-free image.

\textbf{L1 loss:} The pixel-wise mean absolute error between the output and the ground truth image is: 
\begin{equation}
L_{\rm L1}(\hat{\bf I}, {\bf{I}}) = \frac{1}{N} \sum_{n=1}^{N}{\left|\hat{\bf{I}}_n - {\bf I}_n\right|}
\end{equation}

\textbf{L2 loss:} The pixel-wise mean squared error between the output and the ground truth images:
\begin{equation}
L_{\rm L2}(\hat{\bf I}, {\bf{I}}) = \frac{1}{N} \sum_{n=1}^{N}{\left(\hat{\bf{I}}_n - {\bf I}_n\right)^2}
\end{equation}

\textbf{SSIM loss:} The SSIM metric comprises of three perceptual components, namely luminance $l(\hat{\bf I}, {\bf{I}})$, contrast $c(\hat{\bf I}, {\bf{I}})$, and structure $s(\hat{\bf I}, {\bf{I}})$, as shown below.
\begin{align}
    {\rm SSIM}(\hat{\bf I}, {\bf{I}}) = l(\hat{\bf I}, {\bf{I}}) \cdot c(\hat{\bf I}, {\bf{I}}) \cdot s(\hat{\bf I}, {\bf{I}})  \label{eq-ssim}
\end{align}
The detailed expression and further insights into SSIM are available at \cite{wang2004image, rouse2008understanding}. It is remarkable that two images should be similar to each other in terms of the overall luminance, contrast and structure for the SSIM value to be large, which trends at the level of individual pixels are not considered too important. The SSIM values are limited to be between 0 and 1 using ReLU, thus we just subtract the SSIM value from 1 to obtain the SSIM loss function as $L_{\rm SSIM}(\hat{\bf I}, {\bf{I}}) = 1 - {\rm SSIM}(\hat{\bf I}, {\bf{I}})$.

\textbf{MS-SSIM loss:} For calculating MS-SSIM \cite{wang2003multiscale}, the image pairs are iteratively scaled down by a factor of 2 down $M$ number of times. Let us denote $\hat{\bf I}_m$ and ${\bf I}_m$ as the denoised and noise-free images after the $m$th scale down. $c(\hat{\bf I}_m, {\bf{I}}_m)$ and $s(\hat{\bf I}_m,
{\bf{I}}_m)$ are calculated for all values of $m\in[1,M]$ while 
$l(\hat{\bf I}_M, {\bf{I}}_M)$ is only calculated only for $M$th scaled down version.
Then, MS-SSIM is computed as:
\begin{equation}
    {\rm MS\mbox{-}SSIM}(\hat{\bf I}, {\bf{I}}) = [l(\hat{\bf I}_M, {\bf{I}}_M)]^{\alpha_{M}} \cdot \prod_{m=1}^{M}{\left(c(\hat{\bf I}_m, {\bf{I}}_m)\right)^{\beta_{m}} \left(s(\hat{\bf I}_m, {\bf{I}}_m)\right)^{\gamma_{m}}}
\end{equation}
\noindent where $\alpha_M$, $\beta_m$, and $\gamma_m$ are powers imparted to luminance, contrast, and structure terms for the relevant scales. In the original article \cite{wang2003multiscale}, their values are set to 1, and we have used the same. 
The MS-SSIM values are also limited to the range $[0,1]$ using ReLU. We therefore subtract the MS-SSIM value from 1 to obtain the MS-SSIM loss function as $L_{\rm MS\mbox{-}SSIM}(\hat{\bf I}, {\bf{I}}) = 1 - {\rm MS\mbox{-}SSIM}(\hat{\bf I}, {\bf{I}})$.

\textbf{Perceptual or VGG loss:} The perceptual loss \cite{} is calculated by comparing the high-level representations obtained by feeding the images to a pretrained benchmark convolutional network, such as VGG-16 \cite{simonyan2014very} (hence the name VGG loss). The activations obtained from the 4th, 9th, 16th and 23rd layer in the VGG-16 model by passing the denoised and noise-free images as inputs are used for comparison. Let ${\hat{\bf A}}_l$ and ${\bf A}_l$ denote the activation maps obtained from the $l$th layer of VGG-16 for the denoised and the noise-free images, i.e. $\hat \bf I$ and $\bf I$, respectively. Then the VGG loss is given as:
\begin{equation}
    L_{\rm VGG}(\hat{\bf I}, {\bf{I}}) = \sum_{l \in \{4,9,16,23\}}{\left| {{\hat{\bf A}}_l - {{\bf A}}_l} \right|}
\end{equation}

\textbf{Weighted combination: }
Apart from the loss functions described above, a few more loss functions were devised by using a weight sum of two loss functions. 
\begin{equation}
    L_{\rm combo}(\hat{\bf I}, {\bf{I}}) = (1 - \beta) L_{i}(\hat{\bf I}, {\bf{I}}) + \beta L_{j}(\hat{\bf I}, {\bf{I}})
\end{equation}

Two such combinations are explored - a combination of MS-SSIM and L1 loss functions (with $\beta=0.6$), and a combination of SSIM and L1 loss functions (with $\beta=0.4$). The combination of MS-SSIM and L1 losses is inspired from \cite{zhao2016loss}, where it was found slightly superior to either one of the losses individually. The latter combination (SSIM and L1 loss) was inspired by the appreciable performance of the individual loss functions on the simulated dataset. The optimal weight parameter $\beta$ was determined empirically. 

\paragraph{Training algorithm:}For training, Adam optimizer was used with a learning rate of 0.001. The models were trained for 60 epochs. PyTorch library was used for designing and training the models. 


\section{Results}
We perform validation of our approach using both simulated and actual experimental data. The results and insights are presented below. 
\subsection{Denoising results on simulated validation dataset} \label{sec:results-validation}
A test set was created comprising of 250 image pairs each of actin filaments, mitochondria, and vesicles using the raw microscopy data simulator discussed in section \ref{sec:raw}. 
\paragraph{Quantitative comparison of different methods} We perform quantitative comparison of different models and loss functions using peak signal-to-noise ratio (PSNR), which is a prominent quantitative metrics used for gauging denoising performance. 
For simplicity, we refer to a combination of a loss function and a model as a method. Therefore, essentially, we compare 21 different denoising methods in Table \ref{tab:metric-values} using PSNR. The separation of the test results for the different structures is done to appreciate if the geometry has a bearing on the achievable denoising. It is noted in Table \ref{tab:metric-values} that UNet (R-50) together with VGG performs the best for vesicles and mitochondria and the second best for actin filaments.

\begin{table}[b]
\caption{Quantitative analysis in terms of PSNR for different combinations of models and loss functions. The method with the best PSNR value is highlighted in bold. Further, the three methods M1-M3 used for qualitative comparison are indicated in underline.}
\small
{
    \begin{tabular}{|l||c|c|c|c|c|c|c||c|}
    \hline
     & \multicolumn{7}{c||}{\textbf{Loss functions}}     &  \\ 
    \cline{2-8}\vspace{-2 mm}
    & \textbf{L2} & \textbf{L1} &
    \textbf{SSIM} & \textbf{MS-SSIM} &
    \textbf{VGG} & \textbf{MS-SSIM} &
    \textbf{SSIM} & \textbf{Best loss}\\
    \textbf{Model} &&&&&& $+$ \textbf{L1} & $+$ \textbf{L1} & \textbf{function}\\
    \hline\hline
    \multicolumn{9}{|c|}{Actin filaments}\\
    \hline
    UNet(R-34) & 38.27          & 37.64 & 35.89 & 36.27 & 37.69 & 37.21 & 37.11 & L2 \\
    UNet(R-50) & \underline{37.94}          & 36.61 & 37.58 & 36.74 & \underline{38.60} & \underline{37.43} & 37.99 & VGG \\
    FPN(R-34)  & \textbf{38.71} & 37.26 & 36.76 & 37.31 & {35.28} & 36.93 & 36.25 & L2 \\
    \hline\vspace{-2mm}
    \textbf{Best} & FPN & UNet & UNet & FPN & UNet & UNet & UNet & \\
    \textbf{Model} & R-34 & R-34 & R-50 & R-34 & R-50 & R-50 & R-50 & \\
    \hline\hline
    \multicolumn{9}{|c|}{Vesicles}\\
    \hline
    UNet(R-34) & 39.29 & 38.96 & 38.66 & 37.62 & 39.78          & 37.89 & 38.63 & VGG \\
    UNet(R-50) & \underline{39.64} & 39.18 & 38.62 & 37.28 & \underline{\textbf{39.91}} & \underline{38.31} & 39.61 & VGG \\
    FPN(R-34)  & 38.75 & 37.62 & 37.36 & 37.78 & {35.36}          & 37.87 & 37.57 & L2 \\
    \hline\vspace{-2mm}
    \textbf{Best} & UNet & UNet & UNet & FPN & UNet & UNet & UNet & \\
    \textbf{Model} & R-50 & R-50 & R-34 & R-34 & R-50 & R-50 & R-50 & \\
    \hline\hline
    \multicolumn{9}{|c|}{Mitochondria}\\
    \hline
    UNet(R-34) & 38.13 & 37.19 & 38.05 & {35.46} & 38.97          & 38.63 & 38.21 & VGG \\
    UNet(R-50) & \underline{36.59} & 36.73 & 39.20 & 38.69 & \underline{\textbf{40.36}} & \underline{37.42} & 39.98 & VGG \\
    FPN(R-34)  & 40.05 & 38.18 & 38.21 & 37.35 & 37.00          & 38.87 & 39.97 & L2 \\
    \hline\vspace{-2mm}
    \textbf{Best} & FPN & FPN & UNet & UNet & UNet & FPN & UNet &  \\
    \textbf{Model} & R-34 & R-34 & R-50 & R-50 & R-50 & R-34 & R-50 &  \\
    \hline
    \end{tabular}}
\label{tab:metric-values}
\end{table}

\begin{figure}
    \centering
    \includegraphics[width=0.9\linewidth]{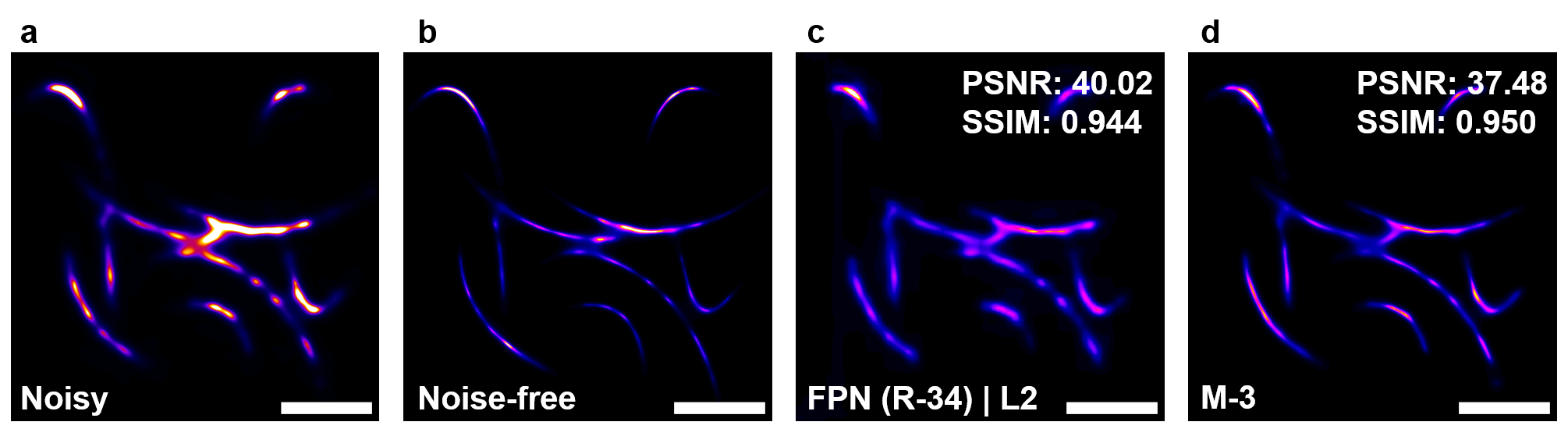}
    \caption{The results of denoising the noisy image (a) using the method with best PSNR (c) and another method M-3 (d) listed in section \ref{sec:results-validation}, and quality comparison with the noise-free image (b). The PSNR and SSIM values for the denoised images are indicated.Scale bar 1 $\upmu$m. }
    \label{fig:discussions}
\end{figure}

\paragraph{Qualitative analysis:} Single valued quantitative metrics such as PSNR are often unsuitable in representing the quality of images, specifically for the case of low-contrast microscopy images. An illustration of this point is given in Fig. \ref{fig:discussions}. The output produced by the method labeled M-3 is much cleaner, with least background debris, while the one produced by FPN (R-34) | L2 has visible artefacts right along the edges of each of the strands. Despite this, the PSNR metric values the former at 37.48 dB while the latter is valued at a much higher PSNR value of 40.02 dB. In contrast, SSIM metric values the denoising output from FPN (R-34) | L2 at a lower score of 0.944 while M-3 is valued at a higher score of 0.95. This is more true to the reality than the PSNR score. However, there will be other cases, where SSIM is not a good indicator of quality. Therefore, we perform a qualitative analysis of aretfact suppression. We consider the following three methods for qualitative comparison:
\begin{itemize}
    \item M-1: UNet (R-50) trained with VGG loss (superior performing in terms of PSNR)\vspace{-3mm}
    \item M-2: UNet(R-50) trained with SSIM + L1 combination loss \vspace{-3mm}
    \item M-3: UNet (R-50) trained with L1 loss
\end{itemize} 

\begin{figure}
    \centering
    \includegraphics[width=\linewidth]{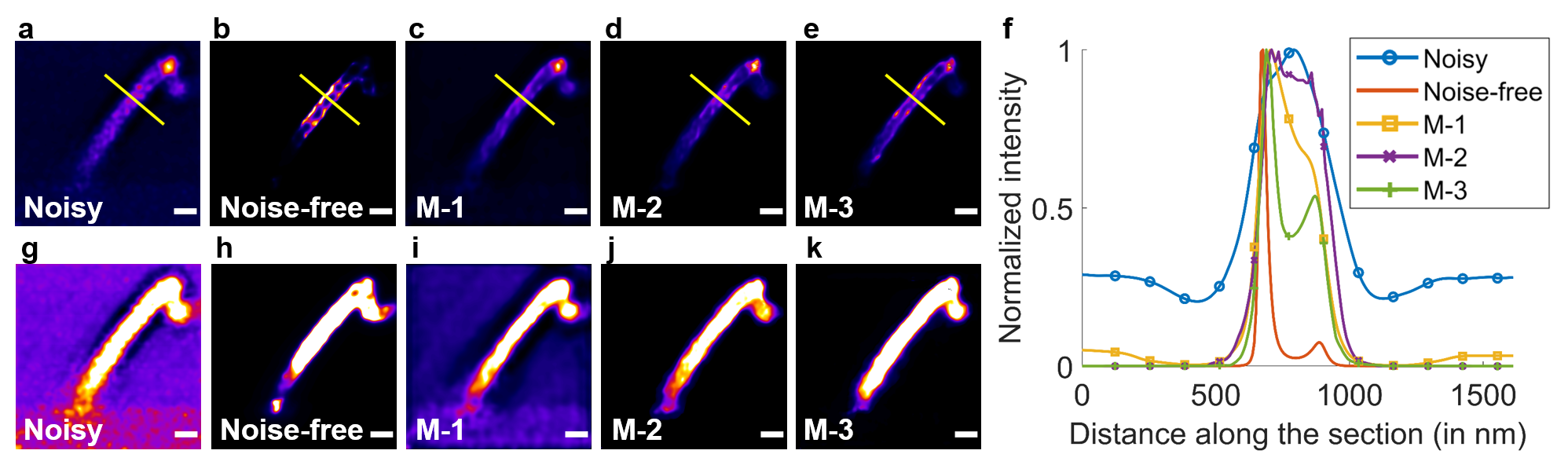}
   \caption{A qualitative comparison for mitochondria where artefact suppression restores resolution. In a-e, the contrast is adjusted manually for best visualization of resolution restoration. The intensities along the yellow line shown in (a-e) are plotted in f. g-k show saturated versions of a-e, where the the out of focus regions and background debris are also visible. Scale bar 500 nm.}
    \label{fig:mitochondria-qualitative}
\end{figure}

The results for mitochondria are presented in Fig. \ref{fig:mitochondria-qualitative}. It is seen that M-1 (\ref{fig:mitochondria-qualitative}c) and M-3 (Fig. \ref{fig:mitochondria-qualitative}e) restore the resolution and construct the boundary of the membrane. M-2 (Fig. \ref{fig:mitochondria-qualitative}d) also appears to perform well, unless the intensities at a line section (shown as yellow line in Fig. \ref{fig:mitochondria-qualitative}a-e) are observed (Fig. \ref{fig:mitochondria-qualitative}f), where M-2 shows a jittery intensity profiles between the two peaks, which may be mistaken as resolving further small features. Here, we have shown only one line section, but we observed similar effect along multiple other sections. Another observation is that the all the methods methods seem to suppress out-of-focus structures (left bottom tail in a-e), but not as effectively as the noise free image (Fig. \ref{fig:mitochondria-qualitative}b). The contrast stretched and over-saturated versions of the images (Fig. \ref{fig:mitochondria-qualitative}g-k) show that the out-of-focus structures are present in all the images, including the noise-free, however with significantly lower intensity as in seen in Fig. \ref{fig:mitochondria-qualitative}b. In this sense, better optical sectioning supported by the noise-free image is still not achieved by the denoised images, although M-3 works the best in this sense. Lastly, from Fig. \ref{fig:mitochondria-qualitative}g-k, it is seen that M-2 and M-3 and significantly more effective in terms of suppressing background debris artefacts. We noted similar observations for actin filaments, i.e. M-3 produces the thinnest filaments and M2-M3 suppress the background debris well. Further, M-3 performs better in suppression of the out-of-focus structures. The results are not reported for space constraints. 

\begin{figure}
    \centering
    \includegraphics[width=\linewidth]{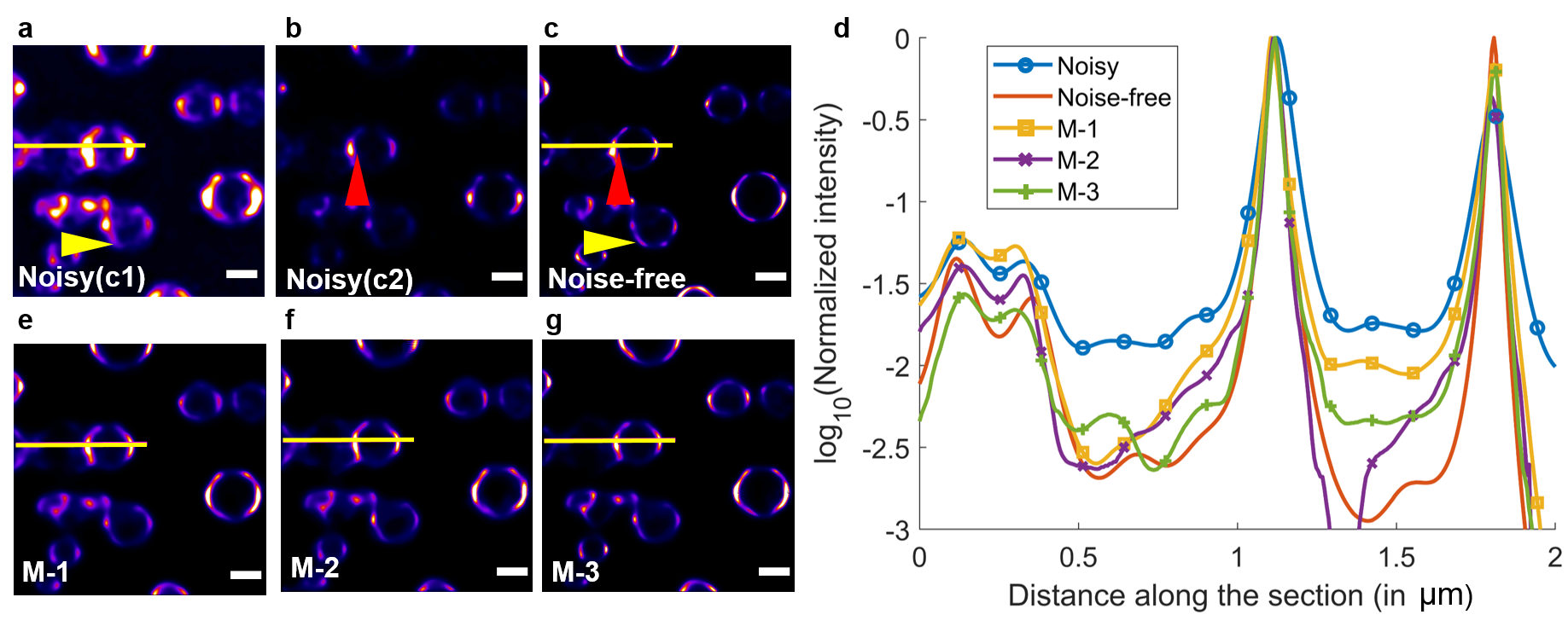}
    \caption{Qualitative comparison for an example of vesicles sample. (a,b) shows the same noise-free image rendered in two contrast stretch. The contrast c1 in (a) is set so that the the structure marked in yellow triangle can be seen in both noisy and noise-free images. The contrast c2 in (b) is set so that the appearance  and visual thickness of the bright spot marked in red triangle appears similar to the noise-free image. The intensities along the line sections shown in (a,b-f) are compared in (g). Scale bar 500 nm.}
    \label{fig:vesicle-qualitative}
\end{figure}

It was indicated in Fig. \ref{fig:ideal-noisy} using red and yellow arrows how the noisy nanoscopy image created background debris due to out-of-focus structures and witnessed reduced sharpness in the features. We show the denoising results for the same example in Fig. \ref{fig:vesicle-qualitative}. The pesudocolor rendering and different contrasts in Fig. \ref{fig:vesicle-qualitative}a-b help in observing these effects more clearly. The yellow line section helps in investigating both the effects simultaneously. The log-intensities at yellow line sections in Fig. \ref{fig:vesicle-qualitative}a,e-g are shown in Fig. \ref{fig:vesicle-qualitative}d. It is seen that the M-1 and M-3 follow quite similar trend with each other and with the noise-free image. Both lower valley in the background region. M-2 generally follows the trend well in the high intensity zones, but may introduce peaks of small intensity in the background.

Overall, M-2 and M-3 are more effective in suppressing background, and M-1 and M-3 are better at improving the sharpness of the image. Generally, for simulated examples, M-3 presents the best qualitative results.

\begin{figure}
    \centering
    \includegraphics[width=\linewidth]{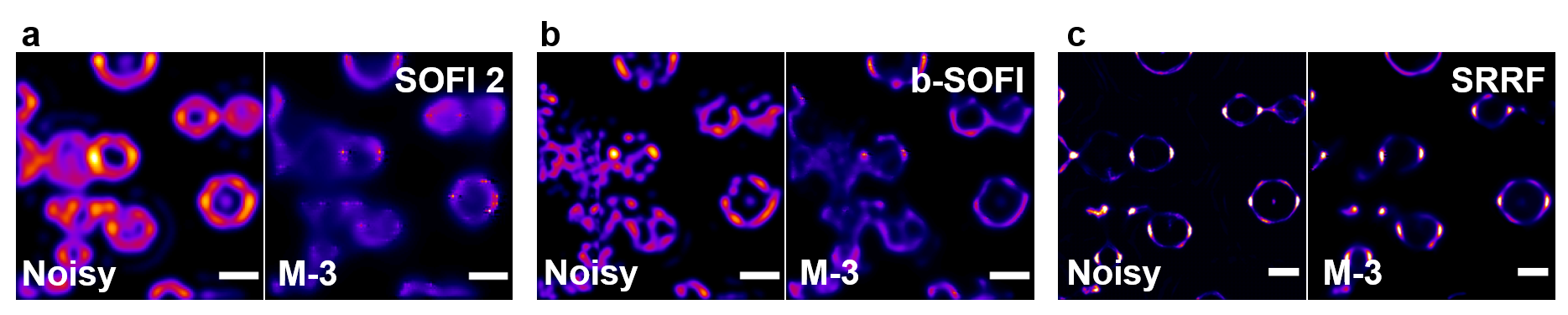}
    \caption{Testing of M-3 method on the nanoscopy images generated by other algorithms. Scale bar 500 nm. \textbf{a.} SOFI order 2. \textbf{b.} bSOFI. \textbf{c.} SRRF with ring radius 2. }
    \label{fig:others}
\end{figure}

\paragraph{Testing nanoscopy images from other nanoscopy algorithms}
Here, we consider if our trained models can be directly applied to nanoscopy images generated by other FNMs. For the same vesicles example as shown in Fig. \ref{fig:ideal-noisy}, we use the noisy raw microscopy image stacks and processed them with three different methods, namely SOFI \cite{dertinger2009fast}, 
bSOFI \cite{Geissbuehler2012}, and super-resolution radial fluctuations \cite{gustafsson2016fast} to obtain noisy nanoscopy images. 
These are then processed using M-3 to generate denoised nanoscopy images. The results are presented in Fig. \ref{fig:others}. It is seen that M-3 does not denoise SOFI and bSOFI images well, but seems to performing well for SRRF. 
Whether it works well on SRRF data of wider variety is still an open question. Therefore, we conclude that even if some transferability may be present across methods that generate similar type of features for certain structures (such as seen here for SRRF on vesicles), such an assumption cannot be generally applicable across FNMs and either fresh training or retraining on data created using specific FNMs should be undertaken. At the same time, we note the concept of the proposed method is generalizable, but not the trained models themselves.

\begin{figure}
    \centering
    \includegraphics[width=0.9\linewidth]{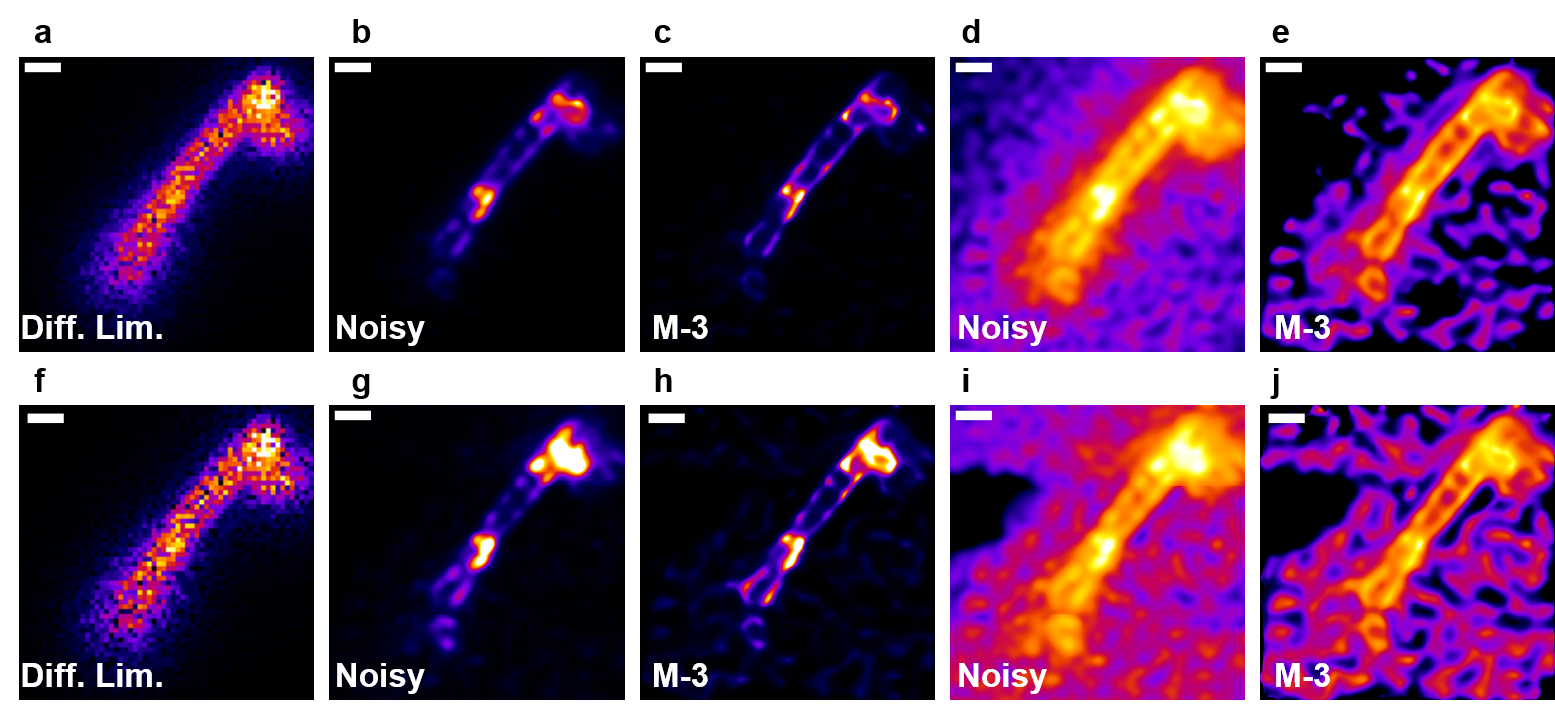}
    \caption{Artefact suppression in nanoscopy image generated using raw microscopy data with speckle noise model. Top row (a-e) and bottom row (f-j) correspond to signal to noise ratio 10 and 5, respectively. (d,e,i,j) correspond to intensity in log scale for (b,c,g,h), respectively.}
    \label{fig:Specklenoise_models}
    \includegraphics[width=0.9\linewidth]{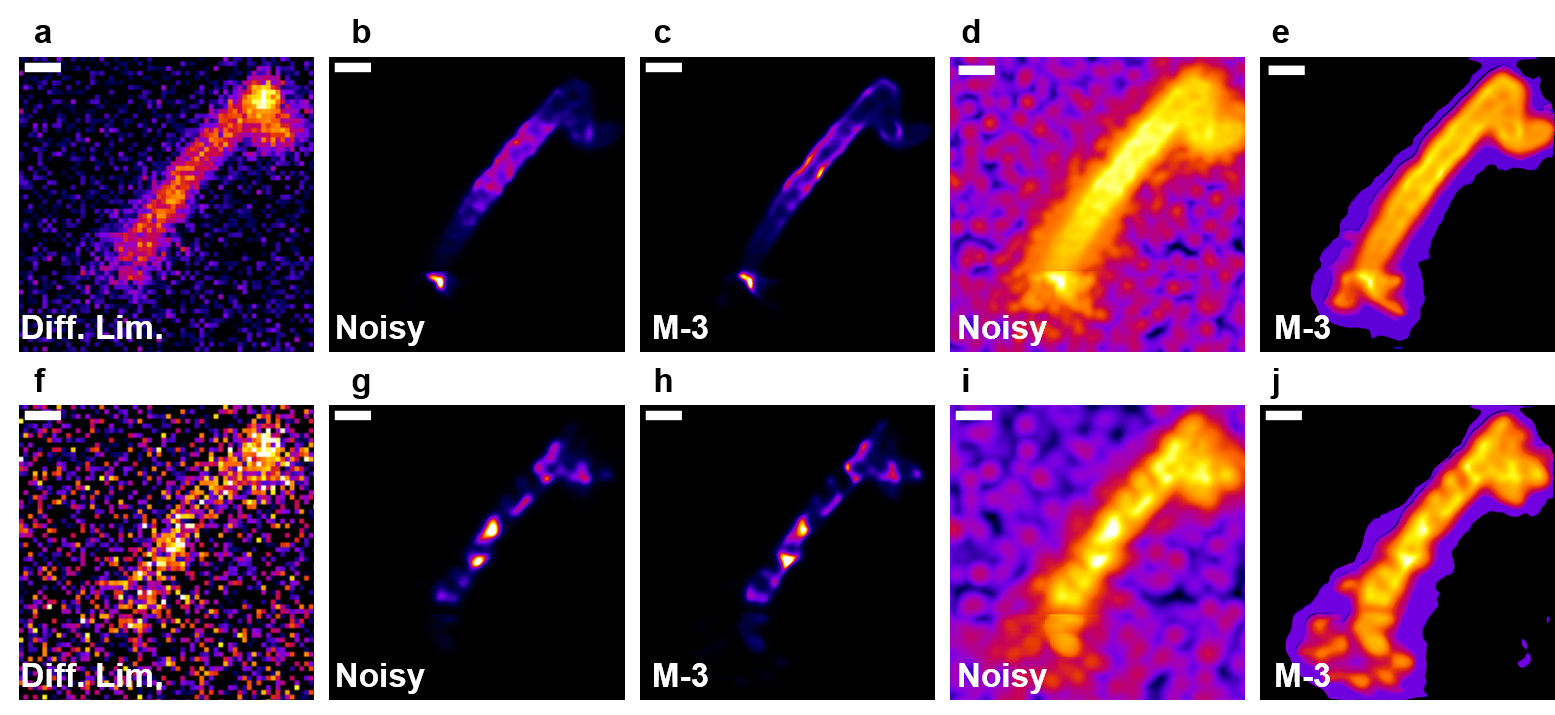}
    \caption{Artefact suppression in nanoscopy image generated using raw microscopy data with Gaussian noise model. Top row (a-e) and bottom row (f-j) correspond to signal to noise ratio 100 and 10, respectively. (d,e,i,j) correspond to intensity in log scale for (b,c,g,h), respectively.}
    \label{fig:Gaussnoise_models}
\end{figure}

\paragraph{Testing nanoscopy images with raw data noise model other than used for simulation}
We conduct this study in order to study the nature of artefacts if noise model than that simulated in raw microscopy data, and assess the generalizability of denoising approach to such noise models in the raw data. We modeled noisy raw data with speckle noise model, assuming variance of 0.1 and 0.2 respectively relative to the max intensity in the noise-free raw data for two different simulations. These correspond to effective signal to noise ratios of 10 and 5, respectively. The results are shown in Fig. \ref{fig:Specklenoise_models}. As expected with the speckle noise model \cite{gonzalez2004digital}, the noise effects the foreground in the raw microscopy data (see Fig. \ref{fig:Specklenoise_models}a,f). Yet, the noisy nanoscopy image can resolve the boundary of the mitochondrion (Fig. \ref{fig:Specklenoise_models}b,g) but contains significant debris in the background (see the log-scale nanoscopy image in Fig. \ref{fig:Specklenoise_models}d,i). We performed denoising using the method M-3. The results indicate that on one hand, denoising makes the boundaries of the mitochondrion sharper (Fig. \ref{fig:Specklenoise_models}c,h), and on the other hand, the debris is suppressed only poorly by M-3 (Fig. \ref{fig:Specklenoise_models}e,j). 

We repeat this experimental with Gaussian noise model. We consider two values of variances, 0.01 and 0.1 relative to the maximum intensity in the noise-free raw data, which corresponds to signal to noise ratio 100 and 10 respectively. The results are presented in Fig. \ref{fig:Gaussnoise_models}. It is seen that the Gaussian noise afflicts the raw microscopy and well as the nanoscopy more severely than the case of speckle noise (Fig. \ref{fig:Specklenoise_models}). Even for signal to noise ratio 100, the resolvability of the boundaries of the mitochondrion is compromised (Fig. \ref{fig:Gaussnoise_models}b) marginally and the background debris visible in the log scale is quite significant in Fig. \ref{fig:Gaussnoise_models}d. Nonetheless, denoising using M-3 restores the boundary of the mitochondrion (Fig. \ref{fig:Gaussnoise_models}c) as well as suppressed the background debris effectively (Fig. \ref{fig:Gaussnoise_models}e). However, in the case of signal to noise ratio 10, the resolvability of the boundaries cannot be restored by denoising (Fig. \ref{fig:Gaussnoise_models}g)  even though the background debris is significantly reduced (Fig. \ref{fig:Gaussnoise_models}j). 

Therefore, it is evident that the different noise distributions in the raw data result into different natures of artefacts and have significant variation in the prominence of artefacts for a given signal to noise ratio. We also note that the denoising method supervised on nanoscopy data generated by raw microscopy with one noise model may be partially effective in reducing artefacts arising from another noise model, for example in terms of resolvability or background debris. In other words, we noted only partial generalizability across raw data noise models.

\subsection{Denoising results on experimental data}
\begin{figure}
    \centering
    \includegraphics[width=0.9\linewidth]{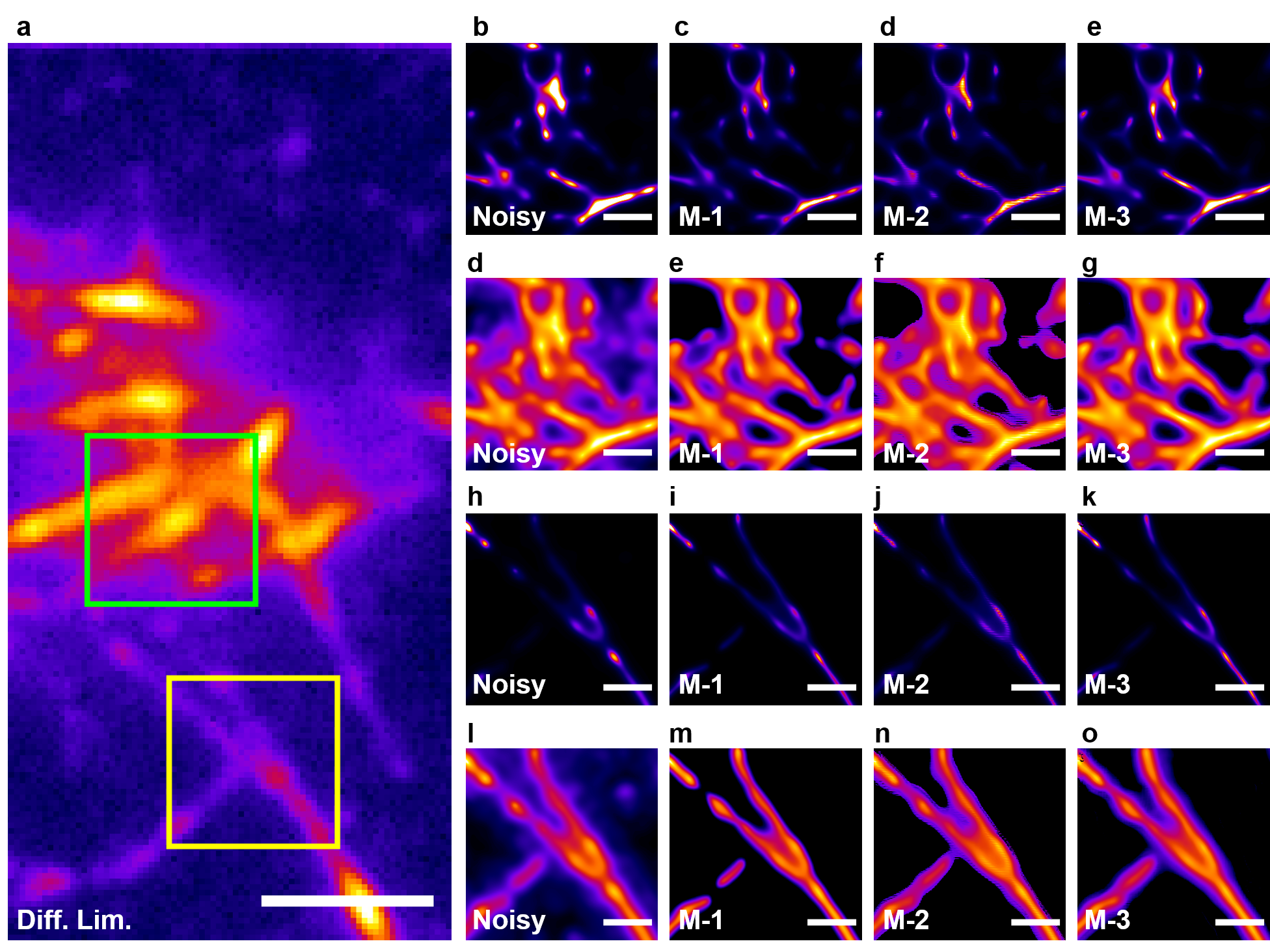}
    \caption{Results of artefact removal from nanoscopy result of in-vitro preformed actin filaments. The second and the fourth rows show results in log scale. The contrast in log scale is adjusted such that the elliptic blob in the top portion of green ROI and the fork in the bottom right of the yellow ROI appear visually similar across the row.  Scale bar 2 $\upmu$m in a, 500 nm in b-o.}
    \label{fig:invitroactin}
\end{figure}

\begin{figure}
    \centering
    \includegraphics[width=0.9\linewidth]{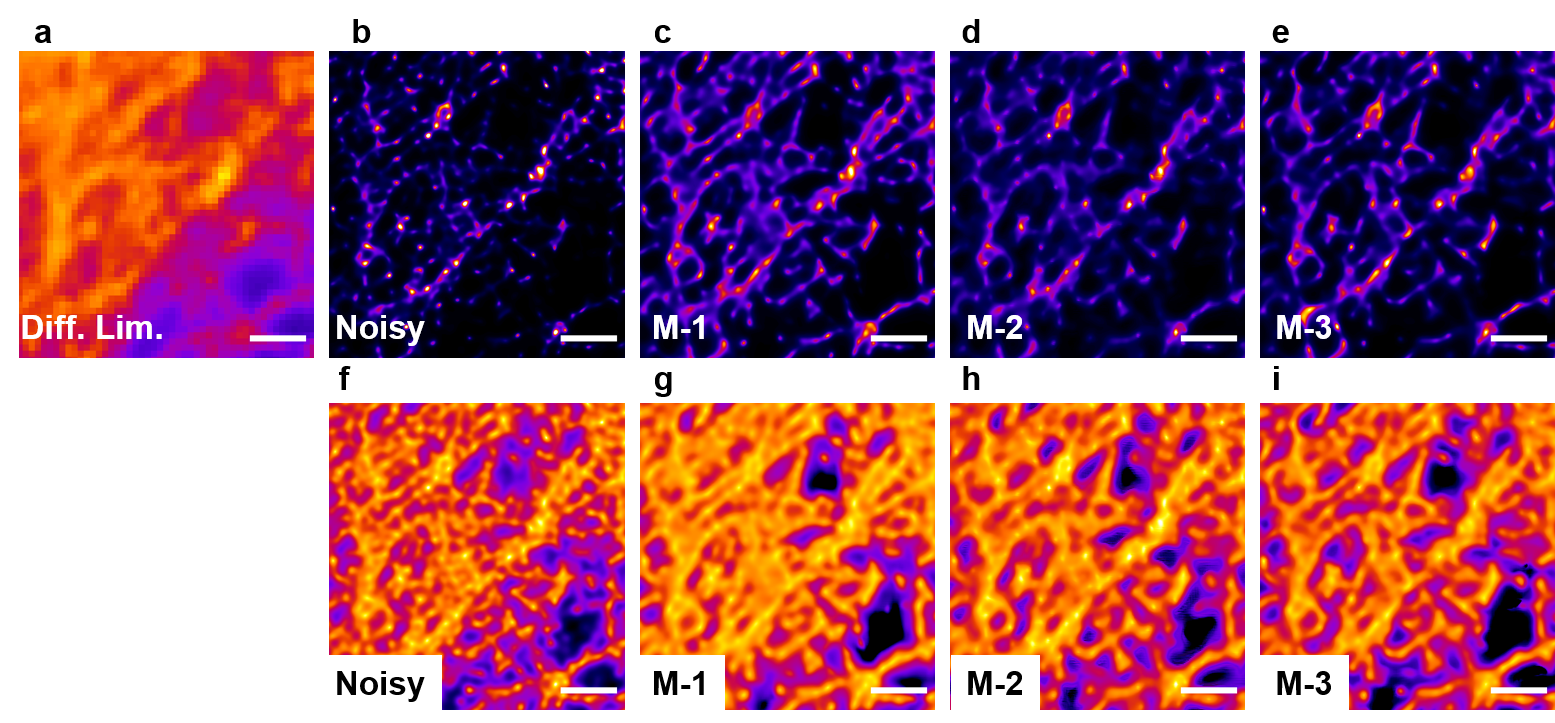}
    \caption{Results of artefact removal from nanoscopy images of microtubules, which were not simulated nor included in the training. Bottom row shows results in the log scale.  Scale bar 1 $\upmu$m.}
    \label{fig:microtubules}
\end{figure}

\begin{figure}
    \centering
    \includegraphics[width=\linewidth]{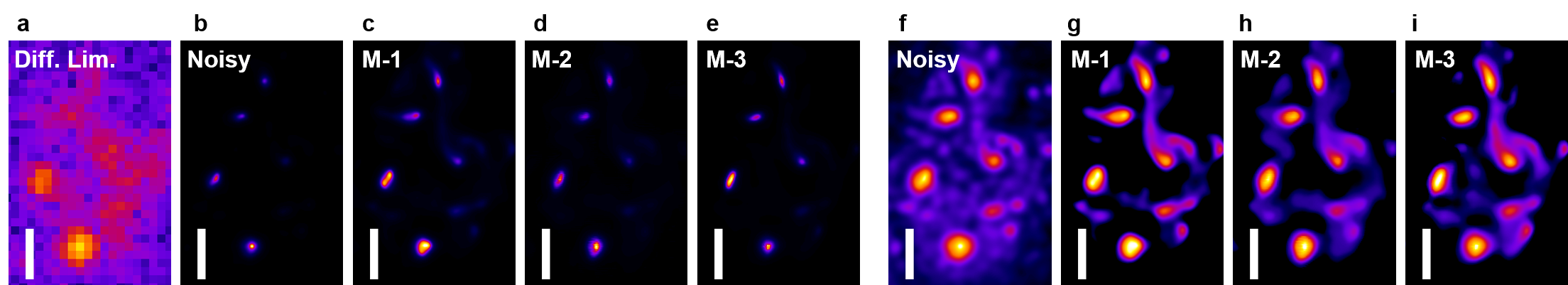}
    \caption{Results of artefact removal from nanoscopy result of liposomes. \textbf{f-i} show results in log scale. The contrast in log scale is adjusted such that the elliptic blob to the left of the color bar appears visually similar. Scale bar 500 nm.}
    \label{fig:liposomes}
\end{figure}

\begin{figure}
    \centering
    \includegraphics[width=\linewidth]{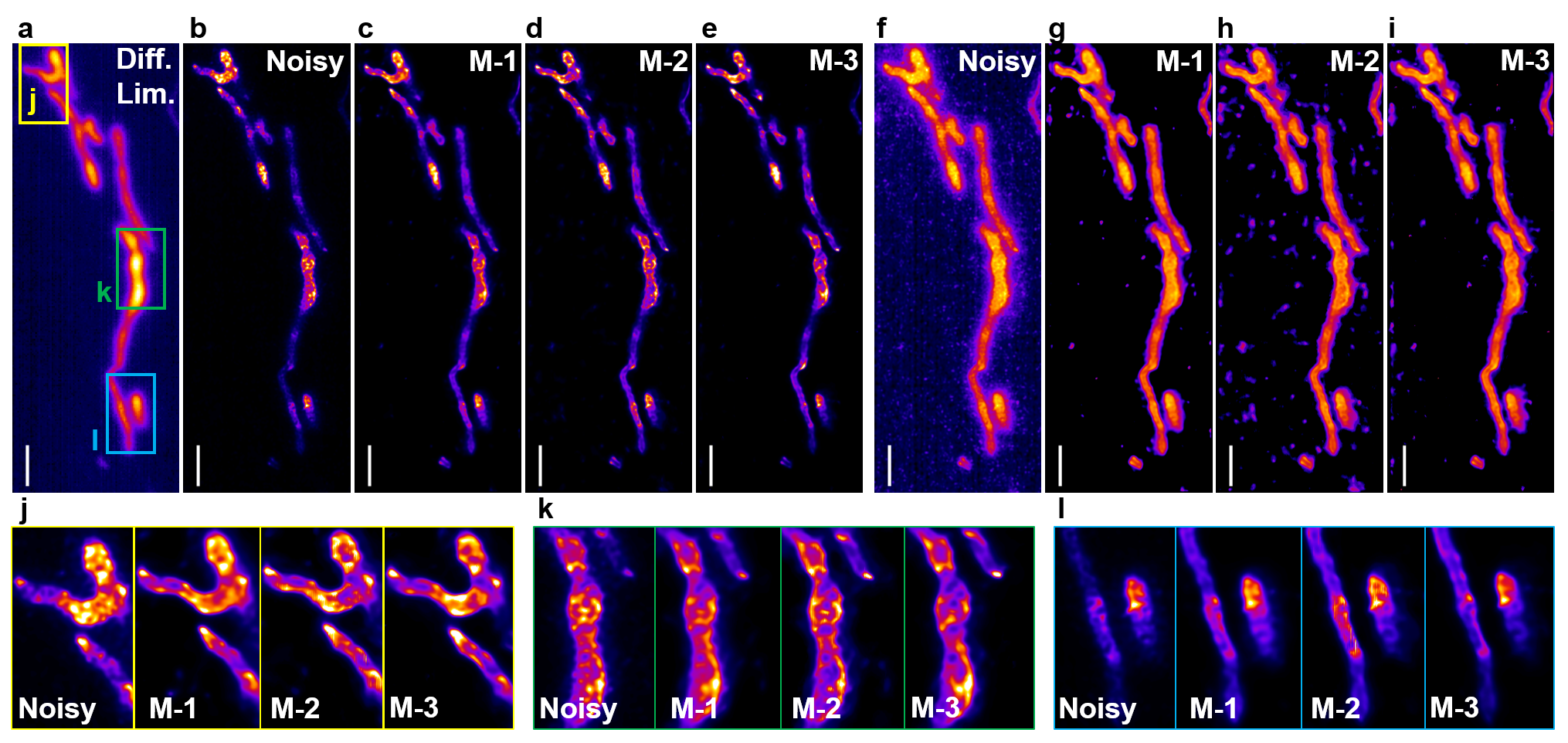}
    \caption{Results of artefact removal from nanoscopy result of mitochondria in living cells. (f-i) show results in log scale. Scale bar 2 $\upmu$m.}
    \label{fig:mitochondria}
\end{figure}

We performed denoising experiments on real microscopy data of actin filaments (invitro preformed), microtubules in fixed cells (these are thick fiber like structures not included in our training data), liposomes (lab-fabricated agarose stablized artificial small vesicles), and mitochondria in living cells. The results for them are presented in Figs. \ref{fig:invitroactin}$-$\ref{fig:mitochondria}, respectively. The experimental details and discussion on results for each data is presented below.  

\paragraph{In vitro preformed actin filaments, Fig. \ref{fig:invitroactin}} \label{sec:invitroactin} This data is taken from the publicly available data of \cite{agarwal2016multiple}. We use the first 500 frames. The relevant imaging parameters are NA 1.49 total internal reflection microscopy, pixel size 65 nm, and emission wavelength of 590 nm. Detailed protocol can be found in \cite{agarwal2016multiple}. The denoising results for a sample of actin filaments are shown in Fig. \ref{fig:invitroactin}. We choose two regions, shown in green and yellow boxes in Fig. \ref{fig:invitroactin}a to consider regions with different local SBRs. The SRBs for the green and yellow boxes are 3.2 and 3.63, despite the peak intensity in the green box being significantly higher. This is because the density of structures in the green box is significantly larger than in the yellow box. We see that all the denoising methods perform similar with minor difference.  It is seen that the portion in the top with a loop that appears saturated in the noisy nanoscopy image (Fig. \ref{fig:invitroactin}b) gets better intensity distributed after denoising (Fig. \ref{fig:invitroactin}c-e). This indicates better contrast distribution close to junctions. The log scale versions of the nanoscopy images for the green box (Fig. \ref{fig:invitroactin}d-g) clearly indicate that the background region is suppressed well by all the methods. However, it is noted that M-3 restores the continuity of some low-intensity strands, a feature that is missed by M-1 and M-2. For the sparser region (yellow box). The denoised results in Fig. \ref{fig:invitroactin}i-k appear similar and are effective in restoring the visibility of the strands. When seen in the log scale (Fig. \ref{fig:invitroactin}l-o), it is evident that M-3 is better at restoring continuity but M-1 is better at suppressing the background faster than M2- and M-3. 

\paragraph{Microtubules in fixed cell, Fig. \ref{fig:microtubules}} \label{sec:microtubules}
We consider an example of microtubules in fixed cells taken from \cite{agarwal2018non} as another challenge case. This is because a microtubule has geometric similarity with actins and mitochondria in the sense of tubularity but is significantly different in terms of radius. The radii of microtubules is in the range 25-30 nm while those of actin filaments are in the range 5-7 nm. Detailed protocol of the considered example can be found in \cite{agarwal2018non}.The first 500 frames of the second example of microtubules in fixed cell are used. This data is also publicly available. The relevant imaging parameters are inverted epifluorescence system of 1.49 NA, 108 nm pixel size, and emission wavelength of 667 nm. The SBR of the selected region is $\sim$4. Since the sample and illumination are 3D, out-of-focus light is also a problem. 

The result is shown in Fig. \ref{fig:microtubules}. The sample has a dense structure with a number of thin strands, not previously encountered in the simulated data. The results are shown in figure \ref{fig:microtubules}. We can see from figures \ref{fig:microtubules} (b - e) that M-1 to M-3 all manage to enhance the continuity of the strands while also suppressing the low intensity, out-of-focus strands. On a qualitative front, M-1 seems to perform the best with good amount of clarity in the individual strands. This example illustrates some potential of generalization of the model for untrained structures and structural density on structures geometrically and optically similar to those simulated for the training dataset.

\paragraph{Liposomes stabilised in agarose, Fig. \ref{fig:liposomes}}
This is one of the challenging samples with structures having radii of 125$\pm$30 nm. This data is also taken from a publicly released dataset of \cite{2008.09195}. The imaging parameters of relevance are epifluorescence microscope of NA 1.42, pixel size 80 nm, and emission wavelength of 537 nm. The liposomes were lab fabricated with average diameter of 250 nm. The emulate vesicles with membrane labels. They were stabilized in agarose, which is likely to contribute to background through autofluorescence. The fragile nature of liposome assembly also means that there may have been debris from liposomes that were disintegrating before the fixation in agarose. These sources of extra background case the SBR at the bright spot seen in Fig. \ref{fig:liposomes}a to be $\sim$4.7 and at the second bright spot to be $\sim$3.2 despite the simplicity of the structures, fixation, and relatively favorable sparsity of liposome distribution. Further, the diffraction limited resolution for the microscope parameters is approximately 190 nm for the noise-free case. Therefore the liposomes comparable in size to the resolution limit. 

The denoising results are shown in Fig. \ref{fig:liposomes}c-d, while Fig. \ref{fig:liposomes}b shows the noisy nanoscopy image. While the denoising or artefact suppression effect is not evident in the denoised images in the first glace, the contrast enhancement and visibility of liposomes other than the two clearly defined ones is witnessed. A further insight is obtained from the log-scaled images shown in Fig. \ref{fig:liposomes}g-k, where the background suppression by M-1 to M-3 is easily noticeable. 

We note that the methods were trained for images with multiple vesicles of radii distributed uniformly in the range [25,500], the radius being selected independently for each vesicle.
Since the intensity of vesicles in the raw microscopy and the nanoscopy images is proportional to the size, smaller object produce dimmer signals and are not trained well for.
This is particularly important for sub-diffraction structures where the resolution-limited image will display intensity proportionally to their sizes.
Since MUSICAL introduces non-linearities in order to achieve super-resolution, this also means that inherently will reduce the contribution and therefore the appearance of dimmer objects.
As a result, the training set is implicitly adding a bias toward larger structure.
However, the structures in the experimental data have a narrow distribution around the resolution limit which explains why the results seems different from the ones obtained for the simulations. Therefore, there exists a margin for customization where the training set contains narrower distribution of the diameter.

\paragraph{Mitochondria in a living cell, Fig. \ref{fig:mitochondria}}
This data is measured in our laboratory on living cardiomyocytes, in which mitochondria were labeled using MitoTracker green dye which are live cell compatible. Two hundred frames were acquired at a frame rate of 40 frames per second but with an exposure time of 3 ms. The other relevant microscopy parameters are epifluorescence microscope of NA 1.42, 80 nm pixel size, and emission wavelength of 520 nm. The SBR of the image stack at the brightest point shown in the green box (labeled k) in Fig. \ref{fig:mitochondria}a is $\sim$3.8. The noisy and denoised nanoscopy images are shown in Fig. \ref{fig:mitochondria}b-e, and their log versions in f-i. The log versions clearly indicate the removal of debris from the background, with the best removal contributed by M-1. In order to facilitate the visualization, zoom-ins of three regions, marked in yellow, green, and blue boxes in Fig. \ref{fig:mitochondria}a are shown in Fig. \ref{fig:mitochondria}j-l. The denoising properties are well exhibited in terms of sharpening the boundaries and suppression of intensities inside the mitochondrial boundaries. Further, it is seen appreciable that the denoising works over regions of different intensities quite well, especially as noted in Fig. \ref{fig:mitochondria}l which corresponds to a region of weak intensity. There, it is notable that all the three methods manage to improve the continuity of the left strand compared to the noisy MUSICAL reconstruction. All the three methods perform equally well on this mitochondria example.

\section{Discussion}
Here we present our observations and comments on various points of interest.  

\paragraph{Generalizability and scalability} Apart from the visually better results obtained on structures that the models were trained on, we observe the models performing generally good on new weakly-related structures that the model was not trained on. We also note a general restoration of resolvability of structures and a reduction in the background debris.
Similar observation extended to noise distributions in the raw data which were not considered in training, where at least partial generalizability of the denoising approach was witnessed. However, since different cameras may have different noise distributions or characteristics, it might be judicious to include a range of noise models in the simulated training dataset.

We deliberately train for an SBR value which is considered quite poor in the hope that it can be generalized for data with better SNR as well. This is clearly witnessed in our results on actual experimental microscopy data. The results also verify that the unconventional approach of simulation-supervised deep learning works well for this problem and helps in circumventing the ground truth absence problem. The random selection of the values of a variety of parameters ensures that diversity of situations are included without introducing significant bias. Nonetheless, some of the quantities at present are fixed either for simplicity of simulations or for limiting the size of dataset (and thereby the time needed for creating it). 
In the future, the same dataset may be expanded for more variety of conditions, or more independent datasets may be created for exploration of transfer learning across FPMs, structures, microscopes; and other sources of artefacts.

\paragraph{Models and loss functions} Coming to specific methods, we see M-1 performing really well on a variety of structures including structures of varying thicknesses, and even densely-packed structures like microtubules. M-2 and M-3 lag slightly behind but still seem to work really well at suppressing the background debris. In summary, M-1 comes across as the most generalizable model producing good results across a variety of structures with appreciable resolution improvement and significant background noise reduction. It is also the method that generally resulted into leading PSNR values in our test data. From our results, it appears that VGG-based perceptual loss function used in M-1 provides good qualitative as well as quantitative performance. It is possibly due to the use activation maps of abstract nature at multiple depths that VGG loss function is able to learn sophisticated artefact suppression model. On the other hand, we think that the combination of SSIM and L1, such as used in M-2, provides a good balance between perceptual quality and pixel wise match.

\paragraph{Metrics and the value of quantitative analysis. }
The training procedure of deep learning methods need loss functions and therefore inherently uses some form of quantitative indicator of quality of denoising. Nonetheless, as exemplified through Table \ref{tab:metric-values} and Fig. \ref{fig:discussions}, a single valued quantitative metric may fail to be an absolute hallmark of quality assessment, especially for the microscopy images in general and nanoscopy images in particular. It might be interesting in the future to design quality metrics customized for this field of science. 

\section{Conclusion}
In this work, artefact removal for a selected fluctuations based nanoscopy method is reported. Artefacts in such nanoscopy methods are attributed to the noise, the photokinetics, as well as the computational treatment of data. A fundamental impediment of the practical artefact removal problem is that it is impossible to experimentally curate a supervised training dataset or synthesize noise-model based datasets. The problem of ground truth absence is effectively dealt with simulations that realistically mimic every aspect of measurement. It is seen that autoencoder deep learning through simulation-supervised training dataset is quite effective in suppressing artefacts arising from photokinetics, raw microscopy, and nanoscopy algorithm induced non-linear data distortions. Our approach is also observed to be generalizable across multiple different structures, different noise models, and nanoscopy algorithms not used during the training process and thus previously unseen by any of the models. Nonetheless, scaling the dataset for more variety of conditions can be easily incorporated or transfer learning can be explored. 
In the future, we wish to add more versatility to the simulation-supervised training dataset and explore the design of suitable metrics for quality analysis in the nanoscopy images.

\section*{Acknowledgements}
The authors acknowledge Ida Sundvor Opstad for generating the data for mitochondria in living cell, Balpreet Singh Ahluwalia for providing the microscope and {\AA}sa Birna Birgisdottir for preparing and providing the cardiomyoblast cells. The providers of the public datasets in section \ref{sec:results-validation} and the teams that contributed to the experimental data in these datasets are also acknowledged.

\section*{Author contributions}
KA and DKP conceived the idea. SJ generated the simulation supervised dataset under the guidance of SA and using the codes provided by him. SJ also performed the deep learning and analysis, with DKP providing insights on the topic. SJ, SA, and KA made the figures. All the authors contributed to writing. 

\section*{Disclosures}
We declare no conflicts of interest. 

\section*{Data availability}
The deep learning models, the simulated raw microscopy data, and the data of mitochondria in living cells will be made public after the manuscript is accepted. The codes will be shared at \url{https://github.com/IAmSuyogJadhav/Nanoscopy-Artefact-Suppresion}.


\bibliography{references}

\end{document}